\documentclass{PoS}
\usepackage{siunitx}
\usepackage[dvipsnames]{xcolor}
\usepackage{comment}

\title{Numerical models of neutrino and gamma-ray emission from 
magnetic reconnection in the core of radio-galaxies}

\ShortTitle{Neutrino and gamma-rays from reconnection in radio-galaxies}

\author{\speaker{J. C. Rodr\'iguez-Ram\'irez}, E. M. de Gouveia Dal Pino \& R. Alves Batista\\
Instituto de Astronomia,  Geof\'isica e Ci\^{e}ncias Atmosf\'ericas (IAG-USP), \\
Universidade de S\~{a}o Paulo \\
R. do Mat\~ao, 1226, Cidade Universit\'aria, 
05508-090,   S\~ao Paulo-SP, Brasil        
\\
E-mail: \email{juan.rodriguez@iag.usp.br}}

\abstract{
Non-blazar radio-galaxies emitting in the very-high-energy (VHE; >100 GeV) regime offer a unique perspective
for probing particle acceleration and emission processes in black hole (BH) accretion-jet systems.
The misaligned nature of these sources indicates the presence of an emission component 
that could be of hadronic origin and located in the core region.
Here we consider turbulent magnetic reconnection in the BH accretion flow of radio-galaxies 
as a potential mechanism for cosmic-ray (CR) acceleration and VHE emission.
To investigate if this scenario is able to account for the observed VHE data, we combine three numerical
techniques to self-consistently model the accretion flow environment and the propagation of CRs plus
electromagnetic cascades within the accretion flow zone.
Here we apply our approach to the radio-galaxy Centaurus A and find that injection of CRs consistent
with magnetic reconnection power partially reproduce the VHE data, provided that the accretion flow makes no
substantial contribution to the radio-GeV components.
The associated neutrino emission peaks at $\sim10^{16}$ eV  and is two
orders of magnitude below the minimum IceCube flux. 
}

\FullConference{International Conference on Black Holes as Cosmic Batteries: UHECRs and Multimessenger Astronomy - BHCB2018\\
		12-15 September, 2018\\
		Foz du Iguazu, Brasil}

\begin{document}
	
\section{Introduction}
Misaligned radio-galaxies emitting at very-high-energies (VHE>100 GeV;
Centaurus A \cite{2018A&A...619A..71H}, 
M87 \cite{2003A&A...403L...1A}, 
IC 310 \cite{2010ApJ...723L.207A}, 
NGC 1275 \cite{2012A&A...539L...2A},
3C 264 \cite{2018ATel11436....1M},
PKS 0625-354 \cite{2018MNRAS.476.4187H}\footnote{The nature of  PKS 0625-354 and IC 310 present features of both radio-galaxies and blazars; see also \cite{2018Galax...6..116R}.
})
offer a unique perspective for probing particle acceleration and emission
processes in black hole (BH) accretion-jet systems. The misaligned nature of these sources
challenge the interpretation of the VHE data  with a 
single zone synchrotron self-Compton (SSC) jet-emission scenario,
as Doppler boosting factors of only a few are allowed
to model the observed spectrum. 
Multi-zone SSC scenarios can explain the
SED including the VHE part, but more free parameters have to be introduced.
(see e.g., \cite{2010A&A...519A..82C}, \cite{2018A&A...619A..71H}).
Larger Doppler factors can be applied 
assuming a bent jet scenario, i.e., that VHE emission is mainly produced within the zone
where the jet is more aligned to the observer line of sight (see e.g., \cite{2014A&A...564A...5A}).
Alternatively, the VHE emission component of radio galaxies could be of is of hadronic 
origin\footnote{Hadronic processes in astrophysics have
recently been favoured by simultaneous neutrino and $\gamma$-ray
detections in the direction of the blazar TXS 0506+056
(\cite{2018Sci...361.1378I})
and two more multi-messenger associations of sources not yet identified \cite{2019ApJ...870..136L}. }
and in the core region 
(\cite{2011A&A...531A..30R}, \cite{2016ApJ...830...81F}, \cite{2014A&A...562A..12P}, \cite{2015arXiv150407592K}, \cite{2016MNRAS.455..838K}).

Here we consider the scenario where the VHE emission of radio-galaxies is produced by the interactions
of CRs with the BH accretion flow, where CRs are accelerated by turbulent magnetic reconnection.
We then investigate the conditions of the accretion flow as well as the distribution
of injected CRs required to reproduce the observed data.
Magnetic reconnection has previously been discussed as an efficient CR acceleration mechanism in turbulent and 
magnetised plasmas (\cite{2005A&A...441..845D}, \cite{2010A&A...518A...5D}, \cite{2011ApJ...735..102K}, \cite{2012PhRvL.108x1102K},
\cite{2015ApJ...799L..20S}, \cite{2018ApJ...864...52K}, \cite{2018MNRAS.481.5687P}).
A description of the phenomenology and analytical and numerical studies of turbulent magnetic reconnection
in BH accretion flows and jets can be found in the contributions by
\cite{2019...PoS...Kadowaki} and \cite{2019...PoS...Gouveia} in these Proceedings. 

In this work, we combine three numerical techniques to model the accretion flow environment and propagation of CRs
to self-consistently investigate the emission and absorption of VHE $\gamma$-rays as well as the production
of neutrinos within the accretion flow zone.
We then compare the CR power required to reproduce the VHE SED with the magnetic reconnection power of the
accretion flow considering the analytical model of \cite{2015ApJ...799L..20S} for  magnetic reconnection in
magnetically dominated accretion flows (MDAF).
Here we focus our analysis on Centaurus A (Cen A), the closest radio-galaxy emitting in the VHE regime. However, our approach 
is aimed to be applied in different sources displaying VHE emission and radiative inefficient accretion flows 
(RIAFs; \cite{2018arXiv181102812R}, \cite{RR_etal_19}).

In the next section we describe the numerical accretion flow simulation that we employ to 
obtain the gas density, magnetic, and photon fields where CRs propagate and interact.
In Section 3 we describe the Monte Carlo simulation of CRs plus electromagnetic cascading 
that we employ to account for gamma-ray and and neutrino emission. 
Finally, we discuss our results in Section 4.

\section{The numerical model for background accretion flow}

We adopt a numerical general relativistic (GR) magneto-hydrodynamic (MHD) RIAF approach together with
GR synchrotron + inverse Compton radiative transfer to model the gas density, magnetic and photon
fields in the BH accretion flow of Cen A. 
To do this, we employ the axi-symmetric
{\tt harm} code \cite{2003ApJ...589..444G}, together with the radiative transfer {\tt grmonty} code 
\cite{2009ApJS..184..387D}.
The accretion flow is simulated within a spatial domain of 40 $R_g$, where the accretion is triggered by
magneto-rotational-instability (MRI) on a torus in initial equilibrium having a poloidal magnetic field with
a maximum plasma beta $\beta=25$. 
We use the dimensionless BH spin parameter $a=0.94$,  
the gas specific heat ratio $\gamma=4/3$ and a 256$\times$256 resolution for the $R$ and $\theta$ spherical coordinates.

Different initial magnetic field configurations are expected to produce different dynamical 
effects in the accretion flow. For instance, a multi-loop poloidal initial magnetic field 
can produce, after enough evolution time, suppression of MRI and very efficient jets, forming the
so-called magnetically arrested disc \cite{2011MNRAS.418L..79T}, \cite{2012MNRAS.423.3083M}).
By contranst, an accretion flow solution with turbulent magnetic field can be 
obtained with an initial magnetic field porfile with a single poloidal loop which produce a less efficient
jet \cite{2018ApJ...853...44O},\cite{2016ApJ...819...95O}. 
For the case of an initial toroidal magnetic field, a spontaneous dipolar magnetic
flux emerge producing transient relativistic jets \cite{2012MNRAS.423.3083M}.
These different flow behaviours are also expected to produce different outcomes in the soft photon
field emission as well as in the CRs emission. We will explore these possibilities in a future work.

For the simulation employed here, we adopt the mass of $M_{BH}=5\times 10^{7}$ M$_\odot$ for the central BH in Cen A to define the spatial and 
temporal scales\footnote{ The spatial and temporal 
scales are defined as $R_g=GM_{BH}/c^2$ and $R_g/c$, respectively, where $G$ is the gravitational
constant, $c$ the speed of light and $M_{BH}$ is the mass of the super massive BH of Cen A.}.
Thus, we define the gas density scale $\rho_0$ fixing the mass accretion rate with the value of 
$\dot{M}_{acc}=1.3\times 10^{-3}$ M$_\odot$ yr$^{-1}\simeq1.17\times 10^{-3}$ M$_{Edd}$ (consistent with the accretion rate obtained by \cite{2009ApJ...695..116E} for Cen A).
Given the gas density scale, the three-magnetic field scale is defined as $B_0=c\sqrt{4\pi\rho_0}$.
\begin{figure}
   \centering
   \includegraphics[width=12cm]{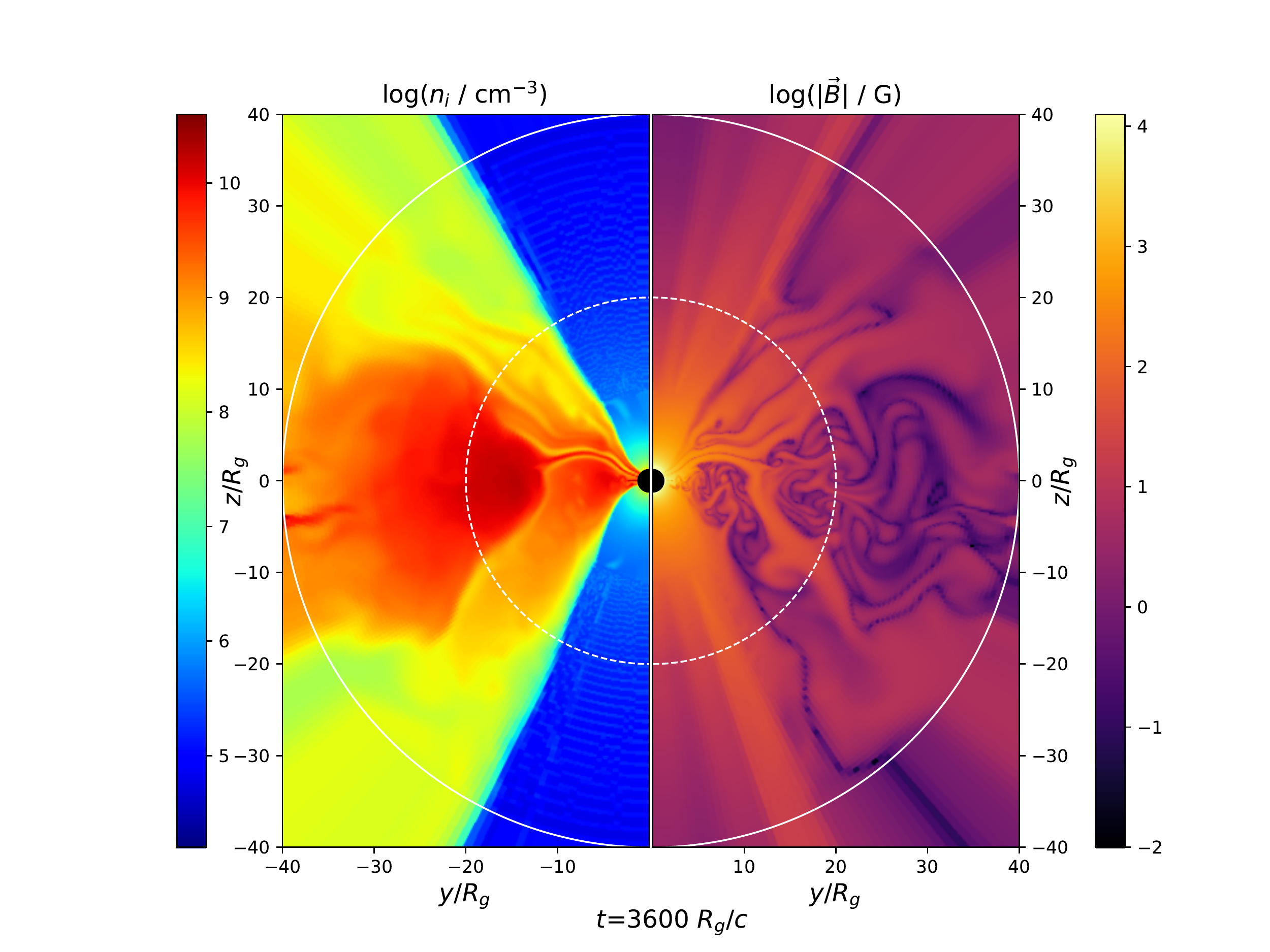}
      \caption{
Gas number density and magnetic field intensity of a simulated accretion flow obtained with the
GRMHD axi-symmetric {\tt harm} code. The accretion rate and the BH mass of Cen A are used to normalise
the physical variables of the snapshot. The inner dashed circle represents the sphere 
of CR injection and the outer solid circle the spherical boundary for particle and photon detections
(see text).
              }
         \label{VHE}
   \end{figure}

For a given snapshot, the photon field is obtained calculating the radiation flux with the 
{\tt grmonty} code at different radius and polar angles within the accretion flow zone. 
Here we assume a constant proton-to-electron temperature ratio $T_p/T_e$ along all the spatial 
domain\footnote{ A more appropriate model for the electron temperature should possibly consider 
its dependence on the plasma magnetisation, see e.g., \cite{2015MNRAS.454.1848R}, \cite{2016A&A...586A..38M},
\cite{2016ApJ...819...95O}.}, where electrons follow a relativistic thermal energy distribution.
In Fig. 1 we show the gas number density and magnetic field maps of the simulation 
snapshot at the integration time of $t=3600 R_g/c$.
We note that in a time interval $\Delta t\simeq 400 R_g/c$  centered in the
snapshot showed in Fig. 1, the accretion flow maintains a turbulent magnetic field, 
with an overall behaviour with no drastic changes. Thus, we choose a sequence
of snapshots within this time interval, to mimic a quasi-continuous injection and emission of
CRs (see the next section).

In Fig. 2 we plot the associated photon field map obtained with $T_p/T_e=190$.
We use the $T_p/T_e$ ratio as a free parameter to obtain different accretion flow models.
These models have the same gas density and magnetic field, but different photon field profiles that
correspond to the values of $T_p/T_e=125$, 190, and 240 (see also Table 1)\footnote{
The temperature of thermal protons is obtained assuming a $\gamma$-law equation
of state $p=(\gamma -1)u$, where p is the gas pressure and $u$ the gas internal energy and thus, $T_p$
$\propto u/\rho$. These variables are given by the GRMHD numerical simulation output and therefore are function of 
the spherical radius $R$ and the polar angle $\theta$.
For the proton-to-electrom temperature values assumed in this work, the electron temperature $T_e$ take values 
in the range of $10^{8-10}$ K, within 20 $R_g$.}
The spectral energy distribution of these photon field profiles are represented in Fig. 4 with 
the histograms in the energy region $<10^{7.5}$ eV.

\begin{figure}
   \centering
   \includegraphics[width=17cm]{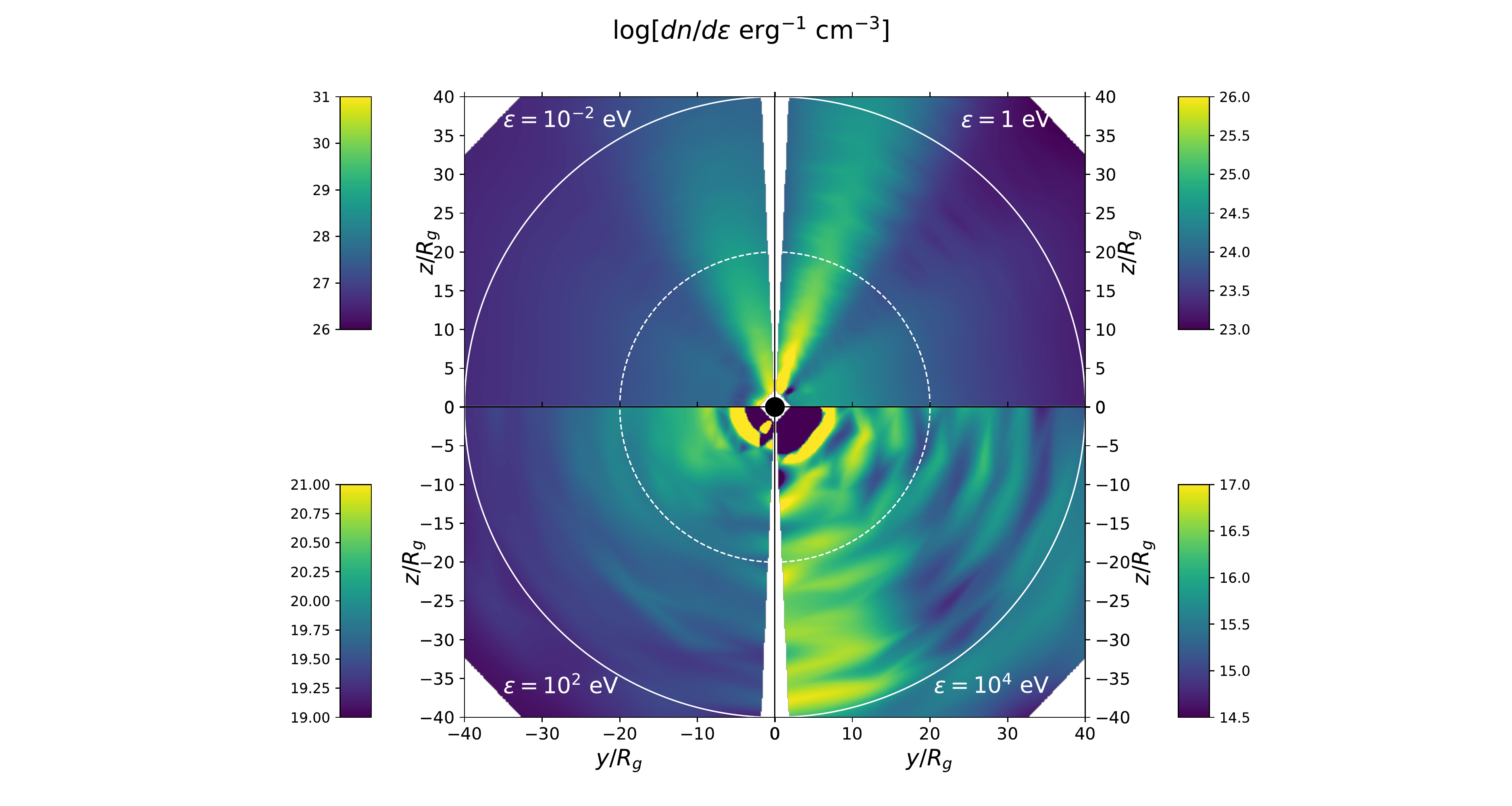}
      \caption{
Photon field maps of the accretion flow snapshot shown in Fig. 1. The maps are obtained performing
radiative transfer of synchrotron and IC radiation with the {\tt grmonty} code.
In each quadrant we plot the photon field density at the specified energy band.}
         \label{VHE}
   \end{figure}

\section{Monte Carlo simulations of CR emission and $\gamma$-$\gamma$+IC cascading}

If magnetic reconnection efficiently accelerates CRs in the accretion flow of Cen A, they could 
potentially produce gamma-rays and neutrinos due to hadronic interactions with the accretion
flow environment. Once they are produced, neutrinos escape practically with no absorption.
On the other hand, $\gamma$-rays are susceptible to be absorbed and develop
electromagnetic cascades due to the soft radiation field of the accretion flow environment. 
Here we assess if the CR power required to reproduce the VHE data of Cen A, can be produced by 
the magnetic reconnection power in the accretion flow.

We consider the magnetic field, gas density and photon field given by the GRMHD model 
described in the previous section, as the target environment for hadronic interactions.
We then employ the Monte Carlo {\tt CRPropa 3} code \cite{2016JCAP...05..038A} to simulate:
(i) proton-proton interactions,
(ii) photomeson interactions,
(iii) $\gamma$-$\gamma$  pair production,
(iv) inverse Compton scattering of secondary leptons produced in the previous interactions, and
(v) synchrotron cooling to account for energy losses of charged particles. However, to save computational resources we do not follow the photons produced by this last process\footnote{
We note that this synchrotron photons have energies < 100 GeV, 
and in this work we are mainly interested in modelling the VHE region of the SED.
Tracking the synchrotron photons to low energies is computationally expensive for
{\tt CRPropa} simulations, where the trajectory of every particle generated within the 
accretion flow zone is followed.}.

The mean-free-paths (MFPs) of particles and photons used in the {\tt CRPropa 3} simulations are calculated
with the gas density and the photon field obtained from the GRMHD simulation.
This accretion flow simulation is  axi-symmetric and thus, the calculated MFPs of particles and photons
vary with the radial and polar positions along their propagation.
We show in Fig. 3 the interaction MFPs corresponding to
the accretion flow snapshot of Fig. 1 and 2 at different radii from the central BH.
In this background environment, CRs with energies $\sim 10^{15-17}$ eV would produce gamma-rays and neutrinos
by photon-pion processes and proton-proton interactions. On the other hand, CRs with energies $\lesssim 10^{14}$ eV
would produce $\gamma$-rays and neutrinos mainly by proton-proton interactions.
The strongest absorption by pair creation occurs for gamma-rays with energies 
of $\sim 10^{14}$ eV. IC scattering of secondary leptons contribute with $\gamma$-ray emission principally
for energies <$10^{13}$ eV.

If the CR trajectories were ballistic within the accretion flow, proton-proton as well as photo-pion
interactions were highly inefficient as the MFPs for this interactions exceed the spatial domain of the 
considered model (40 $R_g$) for most of the energy range (see Fig. 3). 
We note however that CRs are readily trapped by the magnetic field of background environment (see Fig. 1), which favours the efficiency of CR interactions\footnote{
Consider for instance the Larmor radius of a PeV proton in the presence of a magnetic field of $10^2$ G: 
$R_L\simeq 3.3\times 10^6 (E_p$/GeV)(G/$B)$ cm $\simeq 10^{10}$ cm $\simeq 10^{-2} R_g.$}.

\begin{figure}
   \centering
   \includegraphics[width=10cm]{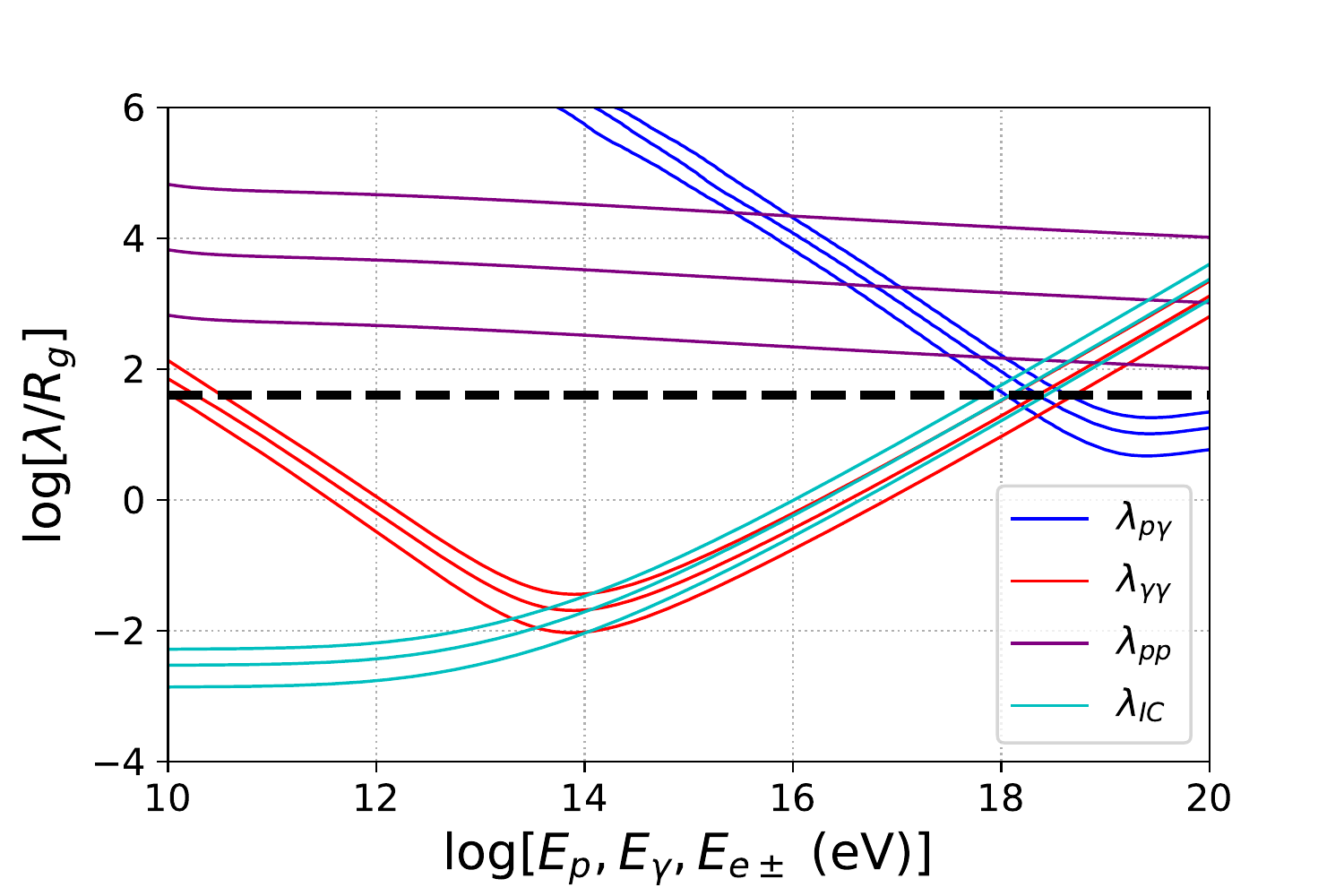}
      \caption{
MFPs for interactions of particles and photons as a function of their energy, employed in the 
simulations performed with the {\tt CRPropa 3} code (see text).
The MFPs are calculated with the gas density and the photon field of the accretion flow snapshot 
shown in Figs. 1 and 2. For the photon-pion ($\lambda_{p\gamma}$), pair production ($\lambda_{\gamma\gamma}$), 
and inverse Compton ($\lambda_{IC}$) interactions, the different curves are
calculated at the $\theta=\ang{45}$ polar angle, and at 20 (lower-most), 30 (middle) and 40 $R_g$ (upper-most) 
from the central BH. 
For the proton-proton interaction ($\lambda_{pp}$)
the different curves are calculated with the maximum gas density (lower-most), and with the 0.1
(middle), and 0.01 (upper-most) fractions of the maximum gas density. 
The black dashed line represents the size of the accretion flow zone (40 $R_g$).  
              }
         \label{VHE}
   \end{figure}

To reproduce the VHE data of Cen A, we consider three accretion flow models, corresponding
to the same mass accretion rate but with different photon field profiles, which are obtained using $T_p/T_e=125$, 190, and 240.
For each model, we emulate a continuous injection of CRs by simulating four burst-like injection of CRs, each one
within the GRMHD snapshots at $t_i=$3440, 3520, 3600, and 3680 $c t/R_g$ (similarly to the CR simulations in \cite{RR_etal_19}). 
We then calculate the observed fluxes of gamma-rays as
$
\nu F_\nu = (4\pi R_s^2)^{-1}\epsilon^2 \dot{N}_\epsilon/\Delta\epsilon
\,\,(\mbox{erg}\, \mbox{s}^{-1}\, \mbox{cm}^{-2})$,
where $R_s=3.8$ Mpc is the distance from CenA to us, $\Delta \epsilon$ is the size of the energy
bin, and $\dot{N}_\epsilon$ is the number of photons per unit time within the energy bin $\epsilon$
that leave the spherical detection boundary at $40 R_g$. The rate of escaping photons is calculated as 
$
\dot{N}_\epsilon =\frac{1}{\Delta t}
\sum\limits_{i=1}^{4} N_{\epsilon,i},
$
where $N_{\epsilon,i}$
are the number of photons within the energy bin $\epsilon$, arriving at the detection boundary (at $40 R_g$),
and produced by the burst-like simulation of CRs injected within the snapshot at the time $t_i$. 
$\Delta t=320 R_g$ is the time interval during which the
code register the arrivals of particles/photons at the spherical
boundary, produced by all the burst-like CR injection events. 

In all background models, we only consider CR protons for simplicity, injecting them  with a power-law energy distribution with exponential cut-off $dN/d\epsilon\propto\epsilon^{-\kappa}\exp\{-\epsilon/\epsilon_{cut}\}$.
According to the analytical calculations of \cite{2015arXiv150407592K}, 
the acceleration rate by magnetic reconnection is balanced by 
CR energy losses at $\sim3\times10^{17}$ eV for parameters corresponding to  Cen A (see
their Fig.3).
On the other hand, \cite{2019MNRAS.485..163K} show that a similar acceleration mechanism driven by reconnection can accelerate CRs up to 10 PeV in typical active galactic nuclei
hosting hot accretion flows.
We then choose the intermediate value of $\epsilon_{cut}=5\times 10^{16}$ eV for the cut-off energy in our models.
For the minimum energy of CR injection we choose the value of $\epsilon_{min}= 10^{13}$ eV, which gives the
best results to reproduce, or partially reproduce the VHE data.

The calculated SED of these models is shown in Fig. 4 and the associated neutrino emission in Fig. 5.  
The power-law index $\kappa$, the power of injected CRs $W_{CR}$, 
and the power of CRs that escape the
spherical detection boundary 
$W_{esc}$ (at 40$R_g$), are listed Table 1 for each emission model.
In this table we also compare the CR power $W_{CR}$ with the magnetic reconnection
power $W_{rec}$ and with the accretion power $\langle \dot{M}_{acc} \rangle c^2$.
The magnetic reconnection power is calculated as a function of the accretion rate and the proton-to-electron 
temperature of the accretion plasma, following the analytical model of \cite{2015ApJ...799L..20S}:
\begin{equation}
W_{rec}=
1.52 \times 10^{42} f
\left( \frac{\dot{M}_{acc}}{\mbox{M}_\odot\mbox{yr}^{-1}}\right)
\left( \frac{T_p}{T_e}\right) \mbox{erg s}^{-1} ,
\end{equation}
where $f$ is a combination of dimensionless 
factors\footnote{We define $f\equiv A\Gamma^{-1}(11.5\alpha^{10/3}+14.89)^{1/2}$, being $A$ the ratio of the
height to the radius of the magnetic reconnection zone, $\Gamma^{-1}=[1+(v_A/c)^2]^{1/2}$ the 
relativistic correction factor of the Alfven velocity, and $\alpha$ the viscosity parameter. 
See \cite{2015ApJ...799L..20S} for details.} 
and here we adopt the value $f=5$.

\begin{figure}
   \centering
   \includegraphics[width=16cm]{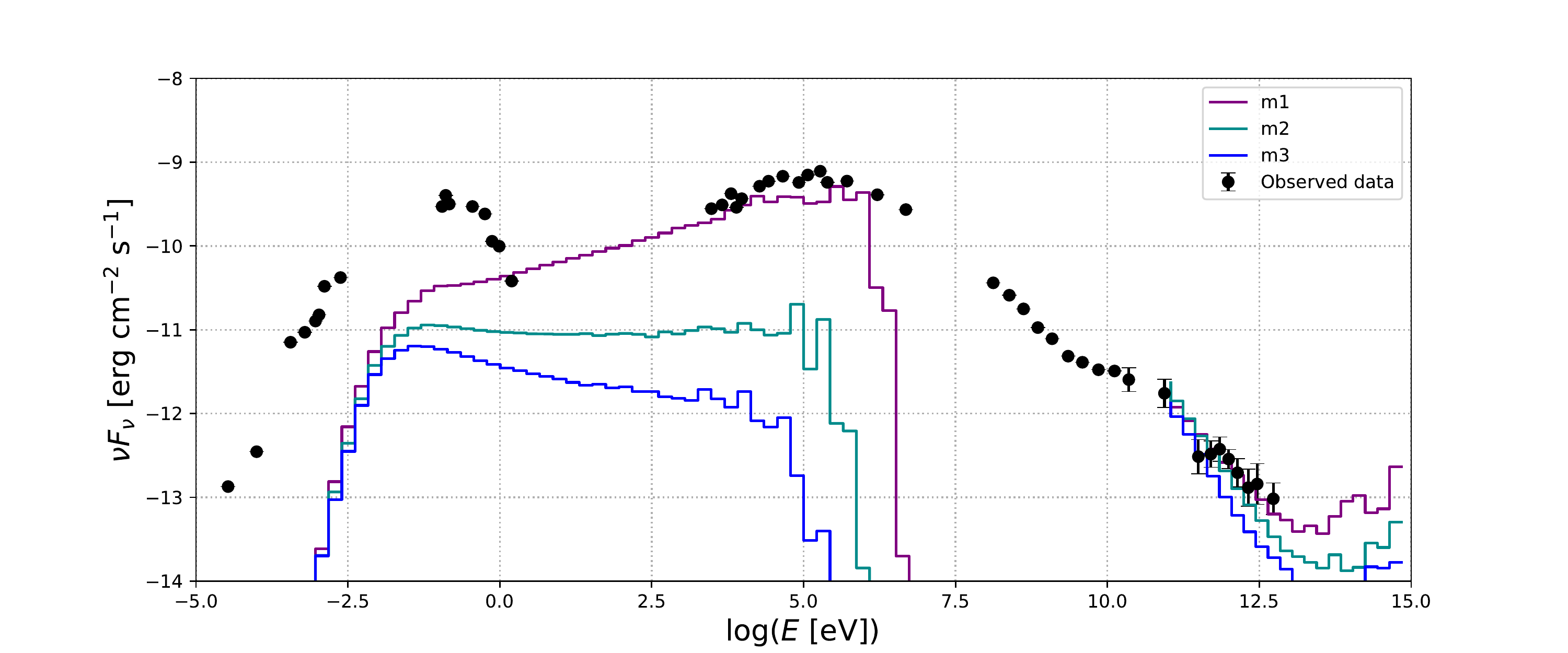}
      \caption{
SED of Cen A. The histograms at energies <$10^{7.5}$ eV are calculated with the {\tt grmonty}
code. The histograms at VHE bands are obtained from the {\tt CRPropa} simulations described in Section 3.
The parameters of the SED models are specified in Table 1. 
The observed points at energies $>10^{7.5}$ eV are taken from \cite{2018A&A...619A..71H}, which are derived from
\textit{Fermi}-LAT and HESS data.
The radio to MeV  data points are adapted from \cite{2018A&A...619A..71H}, which are taken at the same time from
\cite{2010A&A...519A..45O} and \cite{2007A&A...471..453M} (radio to optical), 
\cite{1998A&A...330...97S} (18 keV-8 MeV) and,
\cite{2010PASA...27..431S} (1-30 MeV).
              }
         \label{VHE}
   \end{figure}

\begin{figure}
   \centering
   \includegraphics[width=11cm]{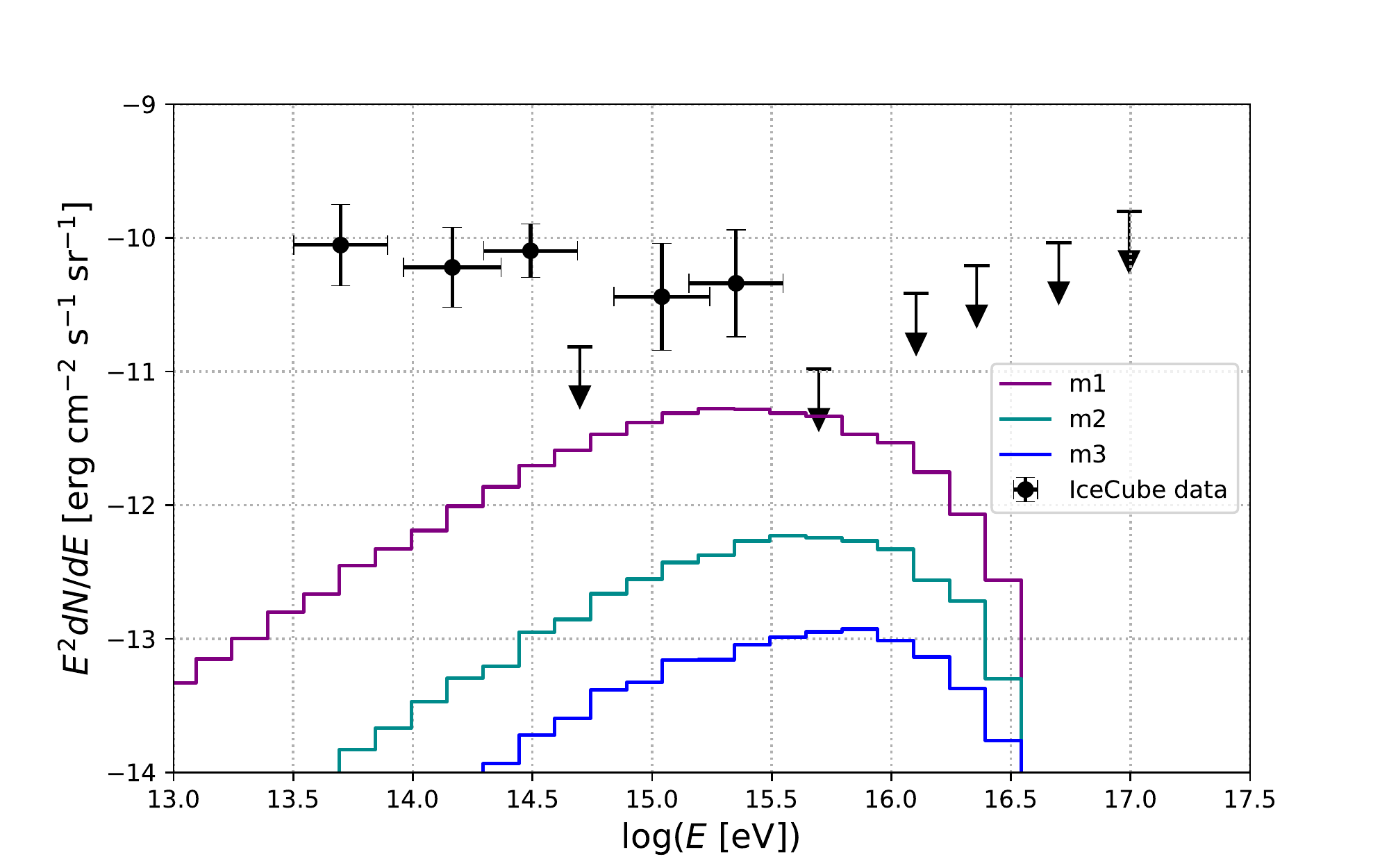}
      \caption{
Neutrino emission models associated with the VHE SED models
for Cen A in Fig. 4. The IceCube data \cite{2014PhRvL.113j1101A} is shown for comparison.
              }
         \label{VHE}
   \end{figure}

\begin{table*}
\caption{
Model parameters of the calculated VHE SED shown in Fig. 4. All the models correspond to a mass 
accretion rate of
$\langle \dot{M}_{acc}\rangle=1.3\times 10^{-3}$ M$_\odot$ yr$^{-1}$ and a CR injection spectrum
$dN_{CR}/d\varepsilon \propto \epsilon^{-\kappa}\exp\{-\varepsilon/\varepsilon_{cut}\}$, with 
$\varepsilon_{min}=10^{13}$ eV and $\varepsilon_{cut}=5\times 10^{16}$ eV.
}
\label{table:1}      
\centering                                      
\begin{tabular}{c c c c c c c c}          
\hline\hline                        
Model  
& $T_p/T_e$ 
& $W_{CR}$ [erg s$^{-1}$]
&$\kappa$ 
& $W_{CR}/W_{rec}$
& $W_{CR}/\langle \dot{M_{acc}}\rangle c^2$
& $W_{esc}$ [erg s$^{-1}$]\\    
\hline                                   
m1 & 125 &   $2\times 10^{43}$  &  1.3 &  13.2  & 2.7$\times 10^{-1}$ & 1.9$\times10^{42}$\\      
m2 & 190 &   $4 \times 10^{42}$ &  1.0 &  2.4 &  5.4$\times 10^{-2}$ & 1.2$\times10^{41}$\\
m3 & 240 &   $9 \times 10^{41}$ &  1.0 &  0.3 &  1.2$\times 10^{-2}$ & 2.6$\times10^{40}$\\
\hline                                             
\end{tabular}
\end{table*}

\section{Summary and Discussion}
We investigate the scenario where the VHE of radio-galaxies is produced by turbulent
magnetic reconnection in the BH accretion flow. To do this, we simulate the interaction of CRs
with the BH accretion flow combining
(i) numerical GRMHD,
(ii) leptonic radiative transfer, and
(iii) Monte Carlo propagation of CR plus electromagnetic cascades within the accretion flow zone
which, based on analytical and numerical results, subject to fast reconnection driven by turbulence 
(see \cite{2019...PoS...Gouveia} and \cite{2019...PoS...Kadowaki} for details).
We then compare the power of CRs required to reproduce the VHE data, with the magnetic reconnection
power given by the analytical model of \cite{2015ApJ...799L..20S}.

Here we apply our approach to model the VHE emission of Cen A and consider three
accretion flow models corresponding to a fixed value of the accretion rate and different values
of the $T_p/T_e$ ratio giving different photon field profiles (see Table 1 and Fig. 4). 
Less power of CR injection is required to match
data points in the VHE regime
for background models with lower photon field densities (see Table 1.). 
This is naturally expected as weaker radiation fields imply less absorption of VHE $\gamma$-rays.
From the models presented in this work, only the model m3 is consistent with the magnetic reconnection
power (i.e., satisfying the condition $W_{CR}/W_{rec}<1$).
This emission model matches only two points of the VHE data and partially contributes to the 
rest of the VHE tail.
The leptonic soft radiation of this model m3 do not contribute substantially to the radio-GeV emission,
which is in agreement with the picture where the radio-GeV data of Cen A is produced in the sub-parsec jet 
by SSC of a single electron population (\cite{2001MNRAS.324L..33C},
\cite{2010ApJ...719.1433A}, \cite{2014ApJ...787...50B}).
Furthermore, the neutrino flux associated to the model m3  
is well below the upper
limits predicted by the IceCube  extragalactic diffuse emission, which is more realistic
for a single source in comparison with the neutrino fluxes of the models m2 and m3.
This is simply because the contribution of photo-pion interactions to neutrino production 
is diminished as the radiation field is lower. We then conclude that CRs accelerated by magnetic 
reconnection are able to partially contribute to the current VHE data 
of Cen A with $\gamma$-rays produced within the accretion flow, provided that this accreting plasma
makes no substantial contribution to the radio-GeV observed emission.

As described in the previous section, the neutrino and VHE radiation fluxes are calculated by simulating
hadronic interactions and electromagnetic cascading inside a spherical boundary at 40 $R_g$ from the BH.
Beyond this detection boundary no further emission and absorption processes
are considered. However, further interactions of the CRs that escape this detection boundary 
might contribute to the observed gamma-ray and neutrinos fluxes. 
Particularly, interactions with the circumnuclear disk (CND) that surrounds the core of Cen A 
(\cite{2008A&A...485L...5M}, \cite{2014A&A...562A..96I}, \cite{2017ApJ...843..136E}),
might appear as core emission if these interactions take place along our line of sight.
The CND is observed with a projected size of $\sim$200$\times$400 pc \cite{2009ApJ...695..116E},
 and has regions with densities of $\sim10^{3-5}$ cm$^{-3}$ (\cite{2017A&A...599A..53I}). 
The mean-free-path of CRs travelling inside the CND could
then be of $\sim 10^{3}$ pc and thus, a non-negligible fraction the CRs escaping the accretion zone
might produce $\gamma$-rays within this gaseous structure.
This propagation effect will be investigated in a forthcoming paper
to complete the study on the VHE emission of Cen A initiated in this proceeding.

\section{Acknowledgments}

We acknowledge support from the Brazilian agencies FAPESP (grant 2013/10559-5) and CNPq (grant 308643/2017-8).
The simulations presented in this lecture have made use of the computing facilities of the GAPAE group (IAG-USP) and the Laboratory of
Astroinformatics IAG/USP, NAT/Unicsul (FAPESP grant 2009/54006-4).
RAB is supported by the FAPESP grant 2017/12828-4 and JCRR by the FAPESP grant 2017/12188-5.

\bibliographystyle{JHEP}
\bibliography{refs}

\providecommand{\href}[2]{#2}\begingroup\raggedright\begin{thebibliography}{10}

\bibitem{2018A&A...619A..71H}
{H.E.S.S.~Collaboration}, H.~{Abdalla}, A.~{Abramowski}, F.~{Aharonian},
  F.~{Ait Benkhali}, E.~O. {Ang{\"u}ner} et~al. \emph{{The {$\gamma$}-ray
  spectrum of the core of Centaurus A as observed with H.E.S.S. and
  Fermi-LAT}}, \href{https://doi.org/10.1051/0004-6361/201832640}{\emph{A\&A}
  {\bfseries 619} A71} (2018)
  [\href{https://arxiv.org/abs/1807.07375}{{\ttfamily 1807.07375}}].

\bibitem{2003A&A...403L...1A}
F.~{Aharonian}, A.~{Akhperjanian}, M.~{Beilicke}, K.~{Bernl{\"o}hr}, H.-G.
  {B{\"o}rst}, H.~{Bojahr} et~al. \emph{{Is the giant radio galaxy M 87 a TeV
  gamma-ray emitter?}},
  \href{https://doi.org/10.1051/0004-6361:20030372}{\emph{A\&A} {\bfseries 403}
  L1} (2003) [\href{https://arxiv.org/abs/astro-ph/0302155}{{\ttfamily
  astro-ph/0302155}}].

\bibitem{2010ApJ...723L.207A}
J.~{Aleksi{\'c}}, L.~A. {Antonelli}, P.~{Antoranz}, M.~{Backes}, J.~A.
  {Barrio}, D.~{Bastieri} et~al. \emph{{Detection of Very High Energy
  {$\gamma$}-ray Emission from the Perseus Cluster Head-Tail Galaxy IC 310 by
  the MAGIC Telescopes}},
  \href{https://doi.org/10.1088/2041-8205/723/2/L207}{\emph{ApJL} {\bfseries
  723} L207} (2010) [\href{https://arxiv.org/abs/1009.2155}{{\ttfamily
  1009.2155}}].

\bibitem{2012A&A...539L...2A}
J.~{Aleksi{\'c}}, E.~A. {Alvarez}, L.~A. {Antonelli}, P.~{Antoranz},
  M.~{Asensio}, M.~{Backes} et~al. \emph{{Detection of very-high energy
  {$\gamma$}-ray emission from NGC 1275 by the MAGIC telescopes}},
  \href{https://doi.org/10.1051/0004-6361/201118668}{\emph{A\&A} {\bfseries
  539} L2} (2012) [\href{https://arxiv.org/abs/1112.3917}{{\ttfamily
  1112.3917}}].

\bibitem{2018ATel11436....1M}
R.~{Mukherjee} \emph{{VERITAS discovery of VHE emission from the FRI radio
  galaxy 3C 264}}, {\emph{The Astronomer's Telegram} {\bfseries 11436} }
  (2018).

\bibitem{2018MNRAS.476.4187H}
{H.E.S.S.~Collaboration}, H.~{Abdalla}, A.~{Abramowski}, F.~{Aharonian},
  F.~{Ait Benkhali}, A.~G. {Akhperjanian} et~al. \emph{{H.E.S.S. discovery of
  very high energy {$\gamma$}-ray emission from PKS 0625-354}},
  \href{https://doi.org/10.1093/mnras/sty439}{\emph{MNRAS} {\bfseries 476}
  4187} (2018) [\href{https://arxiv.org/abs/1802.07611}{{\ttfamily
  1802.07611}}].

\bibitem{2018Galax...6..116R}
F.~{Rieger} and A.~{Levinson} \emph{{Radio Galaxies at VHE Energies}},
  \href{https://doi.org/10.3390/galaxies6040116}{\emph{Galaxies} {\bfseries 6}
  116} (2018) [\href{https://arxiv.org/abs/1810.05409}{{\ttfamily
  1810.05409}}].

\bibitem{2010A&A...519A..82C}
S.~{Colafrancesco}, P.~{Marchegiani} and P.~{Giommi} \emph{{Disentangling the
  gamma-ray emission of NGC 1275 and that of the Perseus cluster}},
  \href{https://doi.org/10.1051/0004-6361/201014393}{\emph{A\&A} {\bfseries
  519} A82} (2010) [\href{https://arxiv.org/abs/1006.2333}{{\ttfamily
  1006.2333}}].

\bibitem{2014A&A...564A...5A}
J.~{Aleksi{\'c}}, S.~{Ansoldi}, L.~A. {Antonelli}, P.~{Antoranz}, A.~{Babic},
  P.~{Bangale} et~al. \emph{{Contemporaneous observations of the radio galaxy
  NGC 1275 from radio to very high energy {$\gamma$}-rays}},
  \href{https://doi.org/10.1051/0004-6361/201322951}{\emph{A\&A} {\bfseries
  564} A5} (2014) [\href{https://arxiv.org/abs/1310.8500}{{\ttfamily
  1310.8500}}].

\bibitem{2018Sci...361.1378I}
{IceCube Collaboration}, M.~G. {Aartsen}, M.~{Ackermann}, J.~{Adams}, J.~A.
  {Aguilar}, M.~{Ahlers} et~al. \emph{{Multimessenger observations of a flaring
  blazar coincident with high-energy neutrino IceCube-170922A}},
  \href{https://doi.org/10.1126/science.aat1378}{\emph{Science} {\bfseries 361}
  1378} (2018) [\href{https://arxiv.org/abs/1807.08816}{{\ttfamily
  1807.08816}}].

\bibitem{2019ApJ...870..136L}
F.~{Lucarelli}, M.~{Tavani}, G.~{Piano}, A.~{Bulgarelli}, I.~{Donnarumma},
  F.~{Verrecchia} et~al. \emph{{AGILE Detection of Gamma-Ray Sources Coincident
  with Cosmic Neutrino Events}},
  \href{https://doi.org/10.3847/1538-4357/aaf1c0}{\emph{ApJ} {\bfseries 870}
  136} (2019) [\href{https://arxiv.org/abs/1811.07689}{{\ttfamily
  1811.07689}}].

\bibitem{2011A&A...531A..30R}
M.~M. {Reynoso}, M.~C. {Medina} and G.~E. {Romero} \emph{{A lepto-hadronic
  model for high-energy emission from FR I radiogalaxies}},
  \href{https://doi.org/10.1051/0004-6361/201014998}{\emph{A\&A} {\bfseries
  531} A30} (2011) [\href{https://arxiv.org/abs/1005.3025}{{\ttfamily
  1005.3025}}].

\bibitem{2016ApJ...830...81F}
N.~{Fraija} and A.~{Marinelli} \emph{{Neutrino, {$\gamma$}-Ray, and Cosmic-Ray
  Fluxes from the Core of the Closest Radio Galaxies}},
  \href{https://doi.org/10.3847/0004-637X/830/2/81}{\emph{ApJ} {\bfseries 830}
  81} (2016) [\href{https://arxiv.org/abs/1607.04633}{{\ttfamily 1607.04633}}].

\bibitem{2014A&A...562A..12P}
M.~{Petropoulou}, E.~{Lefa}, S.~{Dimitrakoudis} and A.~{Mastichiadis}
  \emph{{One-zone synchrotron self-Compton model for the core emission of
  Centaurus A revisited}},
  \href{https://doi.org/10.1051/0004-6361/201322833}{\emph{A\&A} {\bfseries
  562} A12} (2014) [\href{https://arxiv.org/abs/1311.1119}{{\ttfamily
  1311.1119}}].

\bibitem{2015arXiv150407592K}
B.~{Khiali}, E.~M. {de Gouveia Dal Pino} and H.~{Sol} \emph{{Particle
  Acceleration and gamma-ray emission due to magnetic reconnection around the
  core region of radio galaxies}}, {\emph{arXiv e-prints 1504.07592} } (2015)
  [\href{https://arxiv.org/abs/1504.07592}{{\ttfamily 1504.07592}}].

\bibitem{2016MNRAS.455..838K}
B.~{Khiali} and E.~M. {de Gouveia Dal Pino} \emph{{High-energy neutrino
  emission from the core of low luminosity AGNs triggered by magnetic
  reconnection acceleration}},
  \href{https://doi.org/10.1093/mnras/stv2337}{\emph{MNRAS} {\bfseries 455}
  838} (2016) [\href{https://arxiv.org/abs/1506.01063}{{\ttfamily
  1506.01063}}].

\bibitem{2005A&A...441..845D}
E.~M. {de Gouveia Dal Pino} and A.~{Lazarian} \emph{{Production of the large
  scale superluminal ejections of the microquasar GRS 1915+105 by violent
  magnetic reconnection}},
  \href{https://doi.org/10.1051/0004-6361:20042590}{\emph{A\&A} {\bfseries 441}
  845} (2005).

\bibitem{2010A&A...518A...5D}
E.~M. {de Gouveia Dal Pino}, P.~P. {Piovezan} and L.~H.~S. {Kadowaki}
  \emph{{The role of magnetic reconnection on jet/accretion disk systems}},
  \href{https://doi.org/10.1051/0004-6361/200913462}{\emph{A\&A} {\bfseries
  518} A5} (2010) [\href{https://arxiv.org/abs/1005.3067}{{\ttfamily
  1005.3067}}].

\bibitem{2011ApJ...735..102K}
G.~{Kowal}, E.~M. {de Gouveia Dal Pino} and A.~{Lazarian}
  \emph{{Magnetohydrodynamic Simulations of Reconnection and Particle
  Acceleration: Three-dimensional Effects}},
  \href{https://doi.org/10.1088/0004-637X/735/2/102}{\emph{ApJ} {\bfseries 735}
  102} (2011) [\href{https://arxiv.org/abs/1103.2984}{{\ttfamily 1103.2984}}].

\bibitem{2012PhRvL.108x1102K}
G.~{Kowal}, E.~M. {de Gouveia Dal Pino} and A.~{Lazarian} \emph{{Particle
  Acceleration in Turbulence and Weakly Stochastic Reconnection}},
  \href{https://doi.org/10.1103/PhysRevLett.108.241102}{\emph{Physical Review
  Letters} {\bfseries 108} 241102} (2012)
  [\href{https://arxiv.org/abs/1202.5256}{{\ttfamily 1202.5256}}].

\bibitem{2015ApJ...799L..20S}
C.~B. {Singh}, E.~M. {de Gouveia Dal Pino} and L.~H.~S. {Kadowaki} \emph{{On
  the Role of Fast Magnetic Reconnection in Accreting Black Hole Sources}},
  \href{https://doi.org/10.1088/2041-8205/799/2/L20}{\emph{ApJL} {\bfseries
  799} L20} (2015) [\href{https://arxiv.org/abs/1411.0883}{{\ttfamily
  1411.0883}}].

\bibitem{2018ApJ...864...52K}
L.~H.~S. {Kadowaki}, E.~M. {De Gouveia Dal Pino} and J.~M. {Stone} \emph{{MHD
  Instabilities in Accretion Disks and Their Implications in Driving Fast
  Magnetic Reconnection}},
  \href{https://doi.org/10.3847/1538-4357/aad4ff}{\emph{ApJ} {\bfseries 864}
  52} (2018) [\href{https://arxiv.org/abs/1803.08557}{{\ttfamily 1803.08557}}].

\bibitem{2018MNRAS.481.5687P}
M.~{Petropoulou} and L.~{Sironi} \emph{{The steady growth of the high-energy
  spectral cut-off in relativistic magnetic reconnection}},
  \href{https://doi.org/10.1093/mnras/sty2702}{\emph{MNRAS} {\bfseries 481}
  5687} (2018) [\href{https://arxiv.org/abs/1808.00966}{{\ttfamily
  1808.00966}}].

\bibitem{2019...PoS...Kadowaki}
L.~H.~S. {Kadowaki}{\emph{PoS BHCB2018} } (2019).

\bibitem{2019...PoS...Gouveia}
E.~M. {de Gouveia Dal Pino}{\emph{PoS BHCB2018} } (2019).

\bibitem{2018arXiv181102812R}
J.~C. {Rodr{\'{\i}}guez-Ram{\'{\i}}rez}, E.~M. {de Gouveia Dal Pino} and
  R.~{Alves Batista} \emph{{Neutrino and $\gamma$-ray Emission from the Core of
  NGC1275 by Magnetic Reconnection: GRMHD Simulations and Radiative
  Transfer/Particle Calculations}}, {\emph{arXiv e-prints 1811.02812} } (2018)
  [\href{https://arxiv.org/abs/1811.02812}{{\ttfamily 1811.02812}}].

\bibitem{RR_etal_19}
J.~C. {Rodr{\'{\i}}guez-Ram{\'{\i}}rez}, E.~M. {de Gouveia Dal Pino} and
  R.~{Alves Batista} \emph{{VHE emission from magnetic reconnection in the RIAF
  of SgrA*}}, {\emph{Submitted to ApJ} } (2019).

\bibitem{2003ApJ...589..444G}
C.~F. {Gammie}, J.~C. {McKinney} and G.~{T{\'o}th} \emph{{HARM: A Numerical
  Scheme for General Relativistic Magnetohydrodynamics}},
  \href{https://doi.org/10.1086/374594}{\emph{ApJ} {\bfseries 589} 444} (2003)
  [\href{https://arxiv.org/abs/astro-ph/0301509}{{\ttfamily
  astro-ph/0301509}}].

\bibitem{2009ApJS..184..387D}
J.~C. {Dolence}, C.~F. {Gammie}, M.~{Mo{\'s}cibrodzka} and P.~K. {Leung}
  \emph{{grmonty: A Monte Carlo Code for Relativistic Radiative Transport}},
  \href{https://doi.org/10.1088/0067-0049/184/2/387}{\emph{ApJS} {\bfseries
  184} 387} (2009) [\href{https://arxiv.org/abs/0909.0708}{{\ttfamily
  0909.0708}}].

\bibitem{2011MNRAS.418L..79T}
A.~{Tchekhovskoy}, R.~{Narayan} and J.~C. {McKinney} \emph{{Efficient
  generation of jets from magnetically arrested accretion on a rapidly spinning
  black hole}},
  \href{https://doi.org/10.1111/j.1745-3933.2011.01147.x}{\emph{MNRAS}
  {\bfseries 418} L79} (2011)
  [\href{https://arxiv.org/abs/1108.0412}{{\ttfamily 1108.0412}}].

\bibitem{2012MNRAS.423.3083M}
J.~C. {McKinney}, A.~{Tchekhovskoy} and R.~D. {Blandford} \emph{{General
  relativistic magnetohydrodynamic simulations of magnetically choked accretion
  flows around black holes}},
  \href{https://doi.org/10.1111/j.1365-2966.2012.21074.x}{\emph{MNRAS}
  {\bfseries 423} 3083} (2012)
  [\href{https://arxiv.org/abs/1201.4163}{{\ttfamily 1201.4163}}].

\bibitem{2018ApJ...853...44O}
M.~{O' Riordan}, A.~{Pe'er} and J.~C. {McKinney} \emph{{Observational
  Signatures of Mass-loading in Jets Launched by Rotating Black Holes}},
  \href{https://doi.org/10.3847/1538-4357/aaa0c4}{\emph{ApJ} {\bfseries 853}
  44} (2018) [\href{https://arxiv.org/abs/1711.04691}{{\ttfamily 1711.04691}}].

\bibitem{2016ApJ...819...95O}
M.~{O' Riordan}, A.~{Pe'er} and J.~C. {McKinney} \emph{{Jet Signatures in the
  Spectra of Accreting Black Holes}},
  \href{https://doi.org/10.3847/0004-637X/819/2/95}{\emph{ApJ} {\bfseries 819}
  95} (2016) [\href{https://arxiv.org/abs/1510.08860}{{\ttfamily 1510.08860}}].

\bibitem{2009ApJ...695..116E}
D.~{Espada}, S.~{Matsushita}, A.~{Peck}, C.~{Henkel}, D.~{Iono}, F.~P. {Israel}
  et~al. \emph{{Disentangling the Circumnuclear Environs of Centaurus A. I.
  High-Resolution Molecular Gas Imaging}},
  \href{https://doi.org/10.1088/0004-637X/695/1/116}{\emph{ApJ} {\bfseries 695}
  116} (2009) [\href{https://arxiv.org/abs/0901.1656}{{\ttfamily 0901.1656}}].

\bibitem{2015MNRAS.454.1848R}
S.~M. {Ressler}, A.~{Tchekhovskoy}, E.~{Quataert}, M.~{Chandra} and C.~F.
  {Gammie} \emph{{Electron thermodynamics in GRMHD simulations of
  low-luminosity black hole accretion}},
  \href{https://doi.org/10.1093/mnras/stv2084}{\emph{MNRAS} {\bfseries 454}
  1848} (2015) [\href{https://arxiv.org/abs/1509.04717}{{\ttfamily
  1509.04717}}].

\bibitem{2016A&A...586A..38M}
M.~{Mo{\'s}cibrodzka}, H.~{Falcke} and H.~{Shiokawa} \emph{{General
  relativistic magnetohydrodynamical simulations of the jet in M 87}},
  \href{https://doi.org/10.1051/0004-6361/201526630}{\emph{A\&A} {\bfseries
  586} A38} (2016) [\href{https://arxiv.org/abs/1510.07243}{{\ttfamily
  1510.07243}}].

\bibitem{2016JCAP...05..038A}
R.~{Alves Batista}, A.~{Dundovic}, M.~{Erdmann}, K.-H. {Kampert}, D.~{Kuempel},
  G.~{M{\"u}ller} et~al. \emph{{CRPropa 3 -- a public astrophysical simulation
  framework for propagating extraterrestrial ultra-high energy particles}},
  \href{https://doi.org/10.1088/1475-7516/2016/05/038}{\emph{JCAP} {\bfseries
  5} 038} (2016) [\href{https://arxiv.org/abs/1603.07142}{{\ttfamily
  1603.07142}}].

\bibitem{2019MNRAS.485..163K}
S.~S. {Kimura}, K.~{Tomida} and K.~{Murase} \emph{{Acceleration and escape
  processes of high-energy particles in turbulence inside hot accretion
  flows}}, \href{https://doi.org/10.1093/mnras/stz329}{\emph{MNRAS} {\bfseries
  485} 163} (2019) [\href{https://arxiv.org/abs/1812.03901}{{\ttfamily
  1812.03901}}].

\bibitem{2010A&A...519A..45O}
R.~{Ojha}, M.~{Kadler}, M.~{B{\"o}ck}, R.~{Booth}, M.~S. {Dutka}, P.~G.
  {Edwards} et~al. \emph{{TANAMI: tracking active galactic nuclei with austral
  milliarcsecond interferometry . I. First-epoch 8.4 GHz images}},
  \href{https://doi.org/10.1051/0004-6361/200912724}{\emph{A\&A} {\bfseries
  519} A45} (2010) [\href{https://arxiv.org/abs/1005.4432}{{\ttfamily
  1005.4432}}].

\bibitem{2007A&A...471..453M}
K.~{Meisenheimer}, K.~R.~W. {Tristram}, W.~{Jaffe}, F.~{Israel}, N.~{Neumayer},
  D.~{Raban} et~al. \emph{{Resolving the innermost parsec of Centaurus A at
  mid-infrared wavelengths}},
  \href{https://doi.org/10.1051/0004-6361:20066967}{\emph{A\&A} {\bfseries 471}
  453} (2007) [\href{https://arxiv.org/abs/0707.0177}{{\ttfamily 0707.0177}}].

\bibitem{1998A&A...330...97S}
H.~{Steinle}, K.~{Bennett}, H.~{Bloemen}, W.~{Collmar}, R.~{Diehl},
  W.~{Hermsen} et~al. \emph{{COMPTEL observations of Centaurus A at MeV
  energies in the years 1991 to 1995}}, {\emph{A\&A} {\bfseries 330} 97}
  (1998).

\bibitem{2010PASA...27..431S}
H.~{Steinle} \emph{{Centaurus A at Hard X-Rays and Soft Gamma-Rays}},
  \href{https://doi.org/10.1071/AS09070}{\emph{PASA} {\bfseries 27} 431} (2010)
  [\href{https://arxiv.org/abs/0912.2818}{{\ttfamily 0912.2818}}].

\bibitem{2014PhRvL.113j1101A}
M.~G. {Aartsen}, M.~{Ackermann}, J.~{Adams}, J.~A. {Aguilar}, M.~{Ahlers},
  M.~{Ahrens} et~al. \emph{{Observation of High-Energy Astrophysical Neutrinos
  in Three Years of IceCube Data}},
  \href{https://doi.org/10.1103/PhysRevLett.113.101101}{\emph{Physical Review
  Letters} {\bfseries 113} 101101} (2014)
  [\href{https://arxiv.org/abs/1405.5303}{{\ttfamily 1405.5303}}].

\bibitem{2001MNRAS.324L..33C}
M.~{Chiaberge}, A.~{Capetti} and A.~{Celotti} \emph{{The BL Lac heart of
  Centaurus A}},
  \href{https://doi.org/10.1046/j.1365-8711.2001.04642.x}{\emph{MNRAS}
  {\bfseries 324} L33} (2001)
  [\href{https://arxiv.org/abs/astro-ph/0105159}{{\ttfamily
  astro-ph/0105159}}].

\bibitem{2010ApJ...719.1433A}
A.~A. {Abdo}, M.~{Ackermann}, M.~{Ajello}, W.~B. {Atwood}, L.~{Baldini},
  J.~{Ballet} et~al. \emph{{Fermi Large Area Telescope View of the Core of the
  Radio Galaxy Centaurus A}},
  \href{https://doi.org/10.1088/0004-637X/719/2/1433}{\emph{ApJ} {\bfseries
  719} 1433} (2010) [\href{https://arxiv.org/abs/1006.5463}{{\ttfamily
  1006.5463}}].

\bibitem{2014ApJ...787...50B}
M.~J. {Burke}, E.~{Jourdain}, J.-P. {Roques} and D.~A. {Evans} \emph{{The Hard
  X-Ray Continuum of Cen A Observed With INTEGRAL SPI}},
  \href{https://doi.org/10.1088/0004-637X/787/1/50}{\emph{ApJ} {\bfseries 787}
  50} (2014) [\href{https://arxiv.org/abs/1404.2287}{{\ttfamily 1404.2287}}].

\bibitem{2008A&A...485L...5M}
R.~{Morganti}, T.~{Oosterloo}, C.~{Struve} and L.~{Saripalli} \emph{{A
  circumnuclear disk of atomic hydrogen in Centaurus A}},
  \href{https://doi.org/10.1051/0004-6361:200809974}{\emph{A\&A} {\bfseries
  485} L5} (2008) [\href{https://arxiv.org/abs/0805.1627}{{\ttfamily
  0805.1627}}].

\bibitem{2014A&A...562A..96I}
F.~P. {Israel}, R.~{G{\"u}sten}, R.~{Meijerink}, A.~F. {Loenen}, M.~A.
  {Requena-Torres}, J.~{Stutzki} et~al. \emph{{The molecular circumnuclear disk
  (CND) in Centaurus A. A multi-transition CO and [CI] survey with Herschel,
  APEX, JCMT, and SEST}},
  \href{https://doi.org/10.1051/0004-6361/201322780}{\emph{A\&A} {\bfseries
  562} A96} (2014) [\href{https://arxiv.org/abs/1402.0999}{{\ttfamily
  1402.0999}}].

\bibitem{2017ApJ...843..136E}
D.~{Espada}, S.~{Matsushita}, R.~E. {Miura}, F.~P. {Israel}, N.~{Neumayer},
  S.~{Martin} et~al. \emph{{Disentangling the Circumnuclear Environs of
  Centaurus A. III. An Inner Molecular Ring, Nuclear Shocks, and the CO to Warm
  H$_{2}$ Interface}},
  \href{https://doi.org/10.3847/1538-4357/aa78a9}{\emph{A\&A} {\bfseries 843}
  136} (2017) [\href{https://arxiv.org/abs/1706.05762}{{\ttfamily
  1706.05762}}].

\bibitem{2017A&A...599A..53I}
F.~P. {Israel}, R.~{G{\"u}sten}, R.~{Meijerink}, M.~A. {Requena-Torres} and
  J.~{Stutzki} \emph{{The outflow of gas from the Centaurus A circumnuclear
  disk. Atomic spectral line maps from Herschel/PACS and APEX}},
  \href{https://doi.org/10.1051/0004-6361/201629396}{\emph{A\&A} {\bfseries
  599} A53} (2017) [\href{https://arxiv.org/abs/1611.05868}{{\ttfamily
  1611.05868}}].

\end{thebibliography}\endgroup
\end{document}